\documentstyle[12pt,graphicx,natbib]{article}
\newcommand{\bj}{\mbox{b$_{\rm\scriptscriptstyle J}$}}
\def\sun{\hbox{$\odot$}}
\def\aap{A\&A\,  }
\def\apj{ApJ\,  }
\def\apjs{ApJS  }
\def\cjaa{Chinese J. Astron. Astrophys.  }

\def\mnras{MNRAS\,  }
\def\nat{Nature\,  }

\def\pre{Phys. Rev. E   }
\def\physa{Phys. A    }
\def\za{Z. Astrophys.  } 

\begin{document}
\title{
Practical Statistics for
the Voids Between Galaxies}

\author
{L. Zaninetti             \\
Dipartimento di Fisica Generale, \\
           Via Pietro Giuria 1   \\
           10125 Torino, Italy
}

\maketitle
\section*{}
The voids between galaxies are identified with
the volumes of the
Poisson Voronoi tessellation.
Two new survival functions for the apparent
radii of voids are derived.
The sectional normalized area of
the Poisson Voronoi tessellation
 is modelled
by the Kiang function and by the exponential function.
Two  new  survival functions with equivalent
sectional radius are therefore
derived; they represent an alternative
to the survival function of voids between
galaxies  as given by the self-similar
distribution.
The spatial appearance of slices
of the 2dF Galaxy Redshift Survey is simulated.
\\
keywords                    \\
{
methods: statistical ;
cosmology: observations;
(cosmology): large-scale structure of the Universe
}

\section{Introduction}

The statistical analysis  of the voids
between galaxies is a field of research
and the following catalogs have been explored:
the two-degree Field Galaxy Redshift Survey (2dFGRS), see
\cite{Patiri2006} and
\cite{Benda-Beckmann2008},
the Sloan Digital Sky Survey (SDSS), see
\cite{Tikhonov2007},
and the combination of 2dFGRS and SDSS, see \cite{Tinker2008}.
The voids between galaxies are usually modelled
by a survival function in the apparent
radius as given by a modified
exponential distribution,
see (3) in
\cite{Benda-Beckmann2008}
or our Section~\ref{sectionself};
this fact is considered here a standard argument
for comparison.
The concept of Voronoi Diagrams
dates back to the vortex theory
applied to the solar system as developed in the
17th century, see \cite{Descartes1644} and Figure~1
in \cite{Aurenhammer2000}.
The name  is due to the two historical records by \cite{voronoi_1907,voronoi}
and the  applications to
astrophysics beginning with ~\cite{kiang}.
The Voronoi diagrams
represent a model of the
voids between galaxies, see
         \cite{Weygaert1989},
         \cite{pierre1990},
         \cite{barrow1990},
         \cite{coles1991},
         \cite{Weygaert1991a},
         \cite{Weygaert1991b},
         \cite{zaninettig},
         \cite{Ikeuchi1991},
         \cite{Subba1992},
         \cite{Weygaert1994},
         \cite{Goldwirth1995},
         \cite{Weygaert2002},
         \cite{Weygaert2003},
         and
         \cite{Zaninetti2006}.

The  Poisson Voronoi tessellation (PVT) is  a particular case of
the Voronoi tessellation in which the seeds  are generated
independently on the $X$, $Y$  and $Z$ axes in 3D through a
subroutine which returns a random real number taken from a uniform
distribution between 0 and 1. For practical purposes, the
subroutine RAN2  was used, see \cite {press}. On adopting  an
astrophysical point of view, the sectional PVT, $V_p(2,3)$, is
very interesting because it allows a comparison with the voids as
observed in slices of galaxies belonging to different catalogs
such as the CFA2 catalog (\cite{geller}), the 2dfGRS
(\cite{Norberg2002}), the 6dF Galaxy Survey (6dFGS)
(\cite{Jones2004}) or the SDSS (\cite{Einasto2003}). The absence
of (i) a numerical analysis through the survival function
 of normalized
areas in 2D and normalized volumes in 3D of PVT
(ii)
a careful exploration of the statistical properties
of $V_p(2,3)$,
leads to the following
questions being posed.
\begin{itemize}
\item
Is it possible to integrate the usual
probability density functions
(PDFs)
which characterize the main parameters of 2D and 3D PVT
in order to obtain an analytical expression
for the survival function?
\item
Is it possible to model the normalized areas
of $V_p(2,3)$  with the known PDFs?
\item
Can we transform the normalized volumes and areas
into equivalent radius distributions?

\item Can we simulate the observed
      slices of galaxies as given, for example,
      by the
      2dF Galaxy Redshift Survey?

\end {itemize}

In order  to answer these questions,
Section~\ref{secadopted} reports the three
major PDFs used to model the normalized
area/volume of 2D/3D PVT as well as
the results of the fit.

Section \ref{secradius} reports the
apparent distribution in effective radius
of the 3D PVT as well as their associated
survival functions.

Section \ref {sectionself} contains the
self-similar
survival function for voids
between galaxies
as well as the associated PDF.

Section \ref{sectionsectional} reports the fit
of the normalized area distribution
of the sectional PVT with the Kiang
function and the exponential distribution.

Section \ref{seclarge} reports our actual
knowledge of the photometric properties
of galaxies as well as a Voronoi simulation.

It is important to  point out
that the  PVT
is not used as a technique
to identify voids in existing data catalogs,
see      \cite{Ebeling1993},
         \cite{Bernardeau1996},
         \cite{Schaap2000},
         \cite{Marinoni2002},
         \cite{Melnyk2006},
         \cite{Schaap2009}
         and
         \cite{Elyiv2009}.

The PVT formalism is here used conversely: to generate mock
catalogs which are later calibrated by observations. On adopting
the point of view of the statistical distributions is important to
underline that the survival function is here identified with
 the cumulative void size distribution function.
We briefly recall that
the cumulative void size distribution function
relates the number of
voids to their size,
analoguosly to the halo mass function which relates number of dark matter
halos to their mass.

\section{The adopted distributions of
the PVT }

\label{secadopted}

A PDF is the first derivative
of a distribution function (DF)
$F(x)$
with respect to $x$.
In the case where the PDF is known
but the DF is unknown,
the following integral is evaluated
\begin{equation}
F(x) = \int_0^x f(x) dx
\quad .
\end{equation}
As a consequence the survival function (SF)
is
\begin{equation}
SF = 1 - F(x) \quad .
\end{equation}

\subsection{The Kiang function}

The gamma variate $H (x ;c )$  (\cite{kiang})
is
\begin{equation}
 H (x ;c ) = \frac {c} {\Gamma (c)} (cx )^{c-1} \exp(-cx)
\quad ,
\label{kiang}
\end{equation}
where $ 0 \leq x < \infty $, $ c~>0$,
and $\Gamma (c)$ is the gamma function with argument $c$.
The Kiang  PDF has a mean of
\begin{equation}
\mu = 1
\quad ,
\end{equation}
and variance
\begin{equation}
\sigma^2 = \frac{1}{c}
\quad .
\end{equation}
In the case of a 1D PVT,
$c=2$  is an
exact analytical result
and conversely
$c$ is supposed to be 4 or 6
for  2D or  3D  PVTs,
respectively, \cite{kiang}.
The  DF of the Kiang function, DF$_K$,
is
\begin{equation}
DF_K =
1 -{\frac {\Gamma  \left( c,cx \right) }{\Gamma  \left( c \right) }}
\quad ,
\end{equation}
where the incomplete
Gamma  function, $ \Gamma (a,z) $,   is defined by
\begin{equation}
\Gamma (a,z) =
\int _{z}^{\infty }\!{{\rm e}^{-t}}{t}^{a-1}{dt}
\quad .
\end{equation}

The survival function $S_K$  is
\begin{equation}
S_K =
{\frac {\Gamma  \left( c,cx \right) }{\Gamma  \left( c \right) }}
\quad  .
\label{survival_kiang}
\end{equation}

\subsection{Generalized gamma}

The generalized gamma
PDF with three parameters $a,b,c$,
\cite{Hinde1980,Ferenc_2007,Tanemura2003}, is
\begin{equation}
f(x;b,c,d) = c \frac {b^{a/c}} {\Gamma (a/c) } x^{a-1} \exp{(-b
x^c)} \quad . \label{gammag}
\end{equation}
The generalized gamma
 has a mean of
\begin{equation}
\mu = \frac
{
{b}^{-\frac{1}{c} }\Gamma \left( \frac {1+a}{c} \right)
}
{
\Gamma \left( {\frac {a}{c}} \right)
}
\quad ,
\end{equation}
and a variance of
\begin{equation}
\sigma^2 = \frac
{
{b}^{- \frac{2}{c} } \left( +\Gamma \left( {\frac {2+a}{c}} \right)
\Gamma \left( {\frac {a}{c}} \right) - \left( \Gamma \left( {\frac {
1+a}{c}} \right) \right) ^{2} \right)
}
{
\left( \Gamma \left( {\frac {a}{c}} \right) \right) ^{2}
}
\quad .
\end{equation}

The  DF    of the generalized gamma is
\begin{equation}
DF_{GG}  =
1 - \frac { \Gamma  \left( {\frac {a}{c}},b{x}^{c} \right) }
 {
  \Gamma
  \left( {\frac {a}{c}} \right)
}
\quad .
\end{equation}

The  SF of the generalized gamma
is
\begin{equation}
S_{GG} =
\frac { \Gamma  \left( {\frac {a}{c}},b{x}^{c} \right) }
 {
  \Gamma
  \left( {\frac {a}{c}} \right)
}
\quad .
\label{survivalgg}
\end{equation}

\subsection{Ferenc--Neda function }

A new PDF
has been recently introduced, \cite{Ferenc_2007},
in order to model the normalized area/volume
in a 2D/3D PVT
\begin{equation}
FN(x;d) = C \times x^{\frac {3d-1}{2} } \exp{(-(3d+1)x/2)}
\quad ,
\label{rumeni}
\end{equation}
where $C$ is a constant,
\begin{equation}
C =
\frac
{
\sqrt {2}\sqrt {3\,d+1}
}
{
2\,{2}^{3/2\,d} \left( 3\,d+1 \right) ^{-3/2\,d}\Gamma \left( 3/2\,d+
1/2 \right)
}
\quad ,
\end{equation}
and $d(d=1,2,3)$ is the
dimension of the space under consideration.
We will call this
function the  Ferenc--Neda  PDF;
it has a mean of
\begin{equation}
\mu = 1
\quad ,
\end{equation}
and variance
\begin{equation}
\sigma^2 = \frac{2}{3d+1}
\quad .
\end{equation}
The  DF of the Ferenc--Neda function
when $d=2$
is
\begin{eqnarray}
DF_{FN2} =   \nonumber   \\
-{\frac {49}{15}}\,{\frac {\sqrt {2}\sqrt {7}{x}^{5/2}{{\rm e}^{-7/2\,
x}}}{\sqrt {\pi }}}-7/3\,{\frac {\sqrt {2}\sqrt {7}{x}^{3/2}{{\rm e}^{
-7/2\,x}}}{\sqrt {\pi }}}
\nonumber  \\
-{\frac {\sqrt {2}\sqrt {7}\sqrt {x}{{\rm e}^
{-7/2\,x}}}{\sqrt {\pi }}}+
{{\rm erf}\left(1/2\,\sqrt {2}\sqrt {7}\sqrt {x}\right)}
\; ,
\end{eqnarray}
where the error function ${\rm  erf} (x)$ is defined as
\begin{equation}
{\rm erf}(x) =
\int _{0}^{x}\!2\,{\frac {{{\rm e}^{-{t}^{2}}}}{\sqrt {\pi }}}{dt}
\; .
\end{equation}

The  SF of the
Ferenc--Neda function when $d=2$ is
\begin{eqnarray}
S_{FN2} =   \nonumber  \\
1+{\frac {49}{15}}\,{\frac {\sqrt {2}\sqrt {7}{x}^{5/2}{{\rm e}^{-7/2
\,x}}}{\sqrt {\pi }}}+7/3\,{\frac {\sqrt {2}\sqrt {7}{x}^{3/2}{{\rm e}
^{-7/2\,x}}}{\sqrt {\pi }}}
\nonumber  \\
+{\frac {\sqrt {2}\sqrt {7}\sqrt {x}{
{\rm e}^{-7/2\,x}}}{\sqrt {\pi }}}-
{{\rm erf}\left(1/2\,\sqrt {2}\sqrt {7}\sqrt {x}\right)}
\ .
\label{survivalfn2}
\end{eqnarray}
The  DF of the
Ferenc--Neda function
when $d=3$ is
\begin{eqnarray}
DF_{FN3} =\nonumber  \\
1-{{\rm e}^{-5\,x}}-5\,{{\rm e}^{-5\,x}}x
-{\frac {25}{2}}\,{{\rm e}^{-
5\,x}}{x}^{2}
\nonumber  \\
-{\frac {125}{6}}\,{{\rm e}^{-5\,x}}{x}^{3}
-{\frac {625}{
24}}\,{x}^{4}{{\rm e}^{-5\,x}}
\;  .
\end{eqnarray}
The  SF of the
Ferenc--Neda function when $d=3$ is
\begin{eqnarray}
S_{FN3} =
{{\rm e}^{-5\,x}}+5\,{{\rm e}^{-5\,x}}x
\nonumber  \\
+{\frac {25}{2}}\,{{\rm e}^{-5
\,x}}{x}^{2}
+{\frac {125}{6}}\,{{\rm e}^{-5\,x}}{x}^{3}
+{\frac {625}{
24}}\,{x}^{4}{{\rm e}^{-5\,x}}
\quad .
\label{survivalfn3}
\end{eqnarray}

\subsection{Numerical results}

\label{secnumerical}

In the following, we will
model the PVT
in which the seeds  are computed through a random
process.
The  $\chi^2$ is computed
according to the formula
\begin{equation}
\chi^2 = \sum_{i=1}^N \frac { (T_i - O_i)^2} {T_i}
\quad  ,
\label{chisquare}
\end {equation}
where $N$ is the number of bins,
$T_i$ the theoretical value
and
$O_i$ the experimental value.
A first test of the  PDFs presented
in the previous section can be done by
analysing the Voronoi cell normalized area-distribution
in 2D, see Table~\ref{table_data2D}.
 \begin{table}
 \caption[]{
Values of $\chi^2$  for
the cell normalized area-distribution function in 2D;
here $T_i$ are the  theoretical frequencies and
$O_i$ are the  sample frequencies.
Here, we have
25087 Poissonian seeds and $40$ intervals in the histogram.
}
 \label{table_data2D}
 \[
 \begin{array}{lll}
 \hline
PDF ~& parameters  & \chi^2  \\ \noalign{\smallskip}
 \hline
 \noalign{\smallskip}
 H (x ;c ) (Eq. (\ref{kiang}))   &  c=3.55 & 83.48   \\
\noalign{\smallskip}
\hline
f(x;d) (Eq.(\ref{rumeni}))      &  d=2 & 71.83  \\
\noalign{\smallskip}
\hline
G(x;a,b,c) (Eq. (\ref{gammag})) & a=3.15  & 58.9 \\
~                               & b=2.72  & ~ \\
~                               & c=1.13  & ~ \\
 \hline
 \end{array}
 \]
 \end {table}

Table~\ref{table_data3D}
reports the parameters of the PDF
of the volume distribution in 3D.

 \begin{table}
 \caption[]{Values of $\chi^2$  for
cell normalized volume-distribution function in 3D.
Here, $T_i$ are the  theoretical frequencies and
  $O_i$ are the sample frequencies.
Here, we have
21378 Poissonian seeds and $40$ intervals in the histogram.
}
 \label{table_data3D}
 \[
 \begin{array}{lll}
 \hline
PDF ~&  parameters   & \chi^2  \\ \noalign{\smallskip}
 \hline
 \noalign{\smallskip}
 H (x ;c ) ~(Eq. (\ref{kiang}))   &  c=5.53 & 93.86   \\
\noalign{\smallskip}
\hline
f(x;d) ~(Eq. (\ref{rumeni}))           &  d=3 & 134.15  \\
\noalign{\smallskip}
\hline
\noalign{\smallskip}
G(x;a,b,c) ~(Eq. (\ref{gammag}))
& a=4.68      & 58.59 \\
\noalign{\smallskip}
~   &  b=3.87  &  ~   \\
\noalign{\smallskip}
~   &  c=1.18  &  ~   \\
 \hline
 \end{array}
 \]
 \end {table}

\begin{figure}
\begin{center}
\includegraphics[width=7cm]{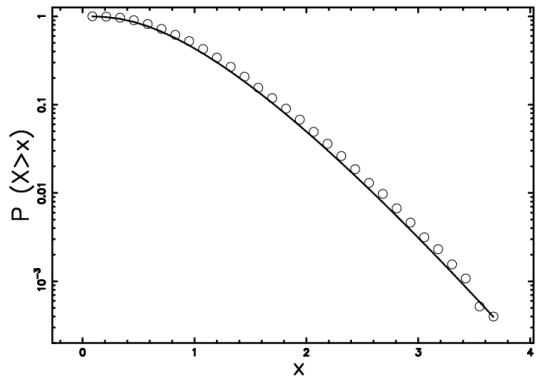}
\end {center}
\caption
{
SF of
normalized area-distribution function in 2D
when we have
25087 Poissonian seeds and $40$ intervals:
the empty circles  represent the Voronoi volumes and
the full line the theoretical value
of $S_{GG}$ (generalized gamma function)
as represented by
(\ref{survivalgg}).
}
 \label{surv_2d_gg}%
 \end{figure}

\begin{figure}
\begin{center}
\includegraphics[width=7cm]{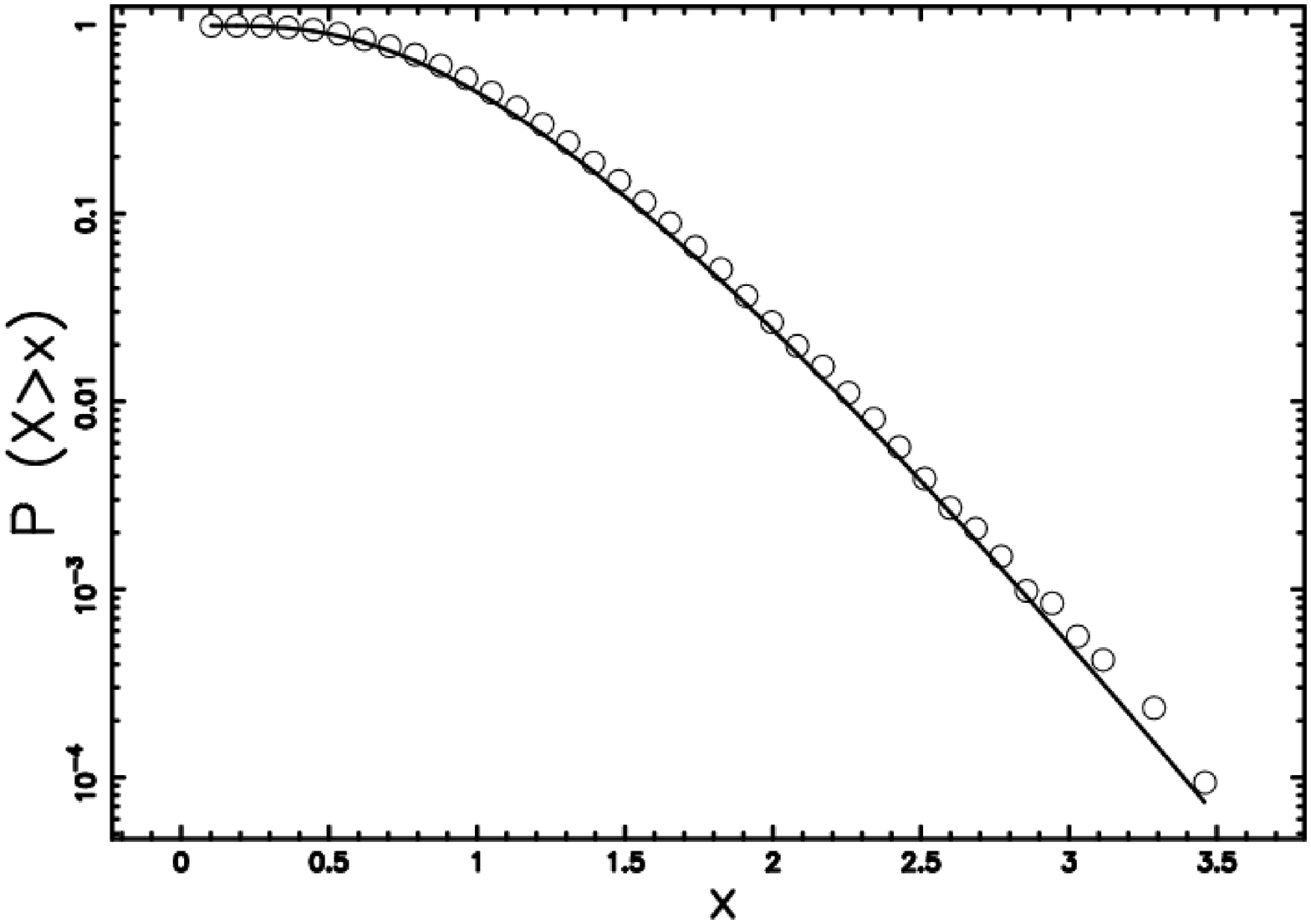}
\end {center}
\caption
{
SF of
normalized volume-distribution function in 3D
when we  have
21378 Poissonian seeds and $40$ intervals:
the empty circles represent the Voronoi volumes and
the full line the theoretical value
of $S_K$ (Kiang function) as represented by
(\ref{survival_kiang}).
}
 \label{surv_3d_kiang}%
 \end{figure}

Figure~\ref{surv_2d_gg} reports an example
of  SF in 2D PVT (areas) and
Figure~\ref{surv_3d_kiang} an
example of SF in 3D PVT (volumes).

\section{Radius distribution of the 3D PVT}

\label{secradius}

We now analyse the distribution in effective radius $R$ of
the 3D PVT.
We assume that the volume of each cell, $v$,
is
\begin{equation}
v = \frac{4}{3} \pi (\frac{R}{\rho})^3 \quad ,
\end{equation}
where $\rho$ is a length  that connects the normalized radius to
the observed one.
In the following, we derive the PDF for radius
and related quantities relative to the Kiang function and
Ferenc--Neda function.

\subsection{Kiang function of the radius }

The PDF, $ H_R (R ;c )$,
of the radius corresponding
to the Kiang function as represented
by (\ref{kiang}) is
\begin{equation}
 H_R (R ;c )   =
\frac {
 4\,c \left( 4/3\,{\frac {c\pi \,{R}^{3}}{{\rho}^{3}}}
\right) ^{c-1}{e ^{-4/3\,{\frac {c\pi \,{R}^{3}}{{\rho}^{3}}}}}\pi
\,{R}^{2} } { \Gamma  \left( c \right) {\rho}^{3} }
 \quad , \label{kiangr}
\end{equation}
where $ 0 \leq R < \infty $, $ c~>0$ and $ \rho ~>0$.
The Kiang PDF of the radius has a mean of
\begin{equation}
\mu = 1/2\,{\frac {\sqrt [3]{2}\sqrt [3]{3}\Gamma  \left( 1/3+c
\right) }{ \sqrt [3]{c}\sqrt [3]{\pi }\Gamma  \left( c \right) }}
\rho \quad ,
\end{equation}
and variance
\begin{eqnarray}
\sigma^2 =  \nonumber  \\
\frac{1}{4}{\frac {{3}^{\frac{2}{3}}{2}^{\frac{2}{3}} \left( \Gamma  \left( 2/3+c \right)
\Gamma  \left( c \right) - \left( \Gamma  \left( 1/3+c \right)
 \right) ^{2} \right) }{{c}^{2/3}{\pi }^{2/3} \left( \Gamma  \left( c
 \right)  \right) ^{2}}}
 \rho^2
\quad .
\end{eqnarray}
The DF of the Kiang function of the radius
is
\begin{equation}
DF_{KR} = 1-{\frac {\Gamma  \left( c,4/3\,c\pi
\,(\frac{R}{\rho})^{3} \right) }{\Gamma
 \left( c \right) }}
\quad .
\end{equation}

The survival function of the radius is
\begin{equation}
S_{KR} = {\frac {\Gamma  \left( c,4/3\,c\pi
\,(\frac{R}{\rho})^{3}  \right) }{\Gamma \left( c \right) }} \quad
. \label{survival_kiangr}
\end{equation}

\subsection{The Ferenc--Neda function of the radius }

The PDF as a function of the radius,
obtained from (\ref{rumeni}) and inserting
$d=3$, is
\begin{equation}
FN_R(R;d) = \frac {400000\,{\pi }^{5}{R}^{14}{e^{-{\frac
{20}{3}}\,{\frac {\pi \,{R}^{3}} {{\rho}^{3}}}}} } {
243\,{\rho}^{15}}
  \quad .
\label{rumenir}
\end{equation}
The mean of the Ferenc--Neda function
is
\begin{equation}
\mu = 0.6  \rho
\quad ,
\end{equation}
and the variance is
\begin{equation}
\sigma^2 = 0.0085 \rho^2\quad .
\end{equation}

The  DF
of the Ferenc--Neda function of the radius
when $d=3$ is
\begin{eqnarray}
DF_{FN3R} =
\nonumber  \\
 1-{e^{-{\frac {20}{3}}\,{\frac {\pi
\,{R}^{3}}{{\rho}^{3}}}}}-{\frac { 20}{3}}\,{e^{-{\frac
{20}{3}}\,{\frac {\pi \,{R}^{3}}{{\rho}^{3}}}}}{R }^{3}\pi
{\rho}^{-3}
\nonumber  \\
-{\frac {200}{9}}\,{e^{-{\frac {20}{3}}\,{\frac { \pi
\,{R}^{3}}{{\rho}^{3}}}}}{R}^{6}{\pi }^{2}{\rho}^{-6} \nonumber \\
-{\frac {4000 }{81}}\,{e^{-{\frac {20}{3}}\,{\frac {\pi
\,{R}^{3}}{{\rho}^{3}}}}}{R} ^{9}{\pi }^{3}{\rho}^{-9}
\nonumber  \\
-{\frac
{20000}{243}}\,{e^{-{\frac {20}{3}}\, {\frac {\pi
\,{R}^{3}}{{\rho}^{3}}}}}{R}^{12}{\pi }^{4}{\rho}^{-12}
\quad .
\end{eqnarray}

The SF of the
Ferenc--Neda function of the radius when $d=3$ is
\begin{eqnarray}
S_{FN3R} =
\nonumber  \\
{e^{-{\frac {20}{3}}\,{\frac {\pi
\,{R}^{3}}{{\rho}^{3}}}}}+{\frac {20 }{3}}\,{e^{-{\frac
{20}{3}}\,{\frac {\pi \,{R}^{3}}{{\rho}^{3}}}}}{R}^ {3}\pi
{\rho}^{-3}
\nonumber  \\
+{\frac {200}{9}}\,{e^{-{\frac {20}{3}}\,{\frac { \pi
\,{R}^{3}}{{\rho}^{3}}}}}{R}^{6}{\pi }^{2}{\rho}^{-6} \nonumber\\
+{\frac {4000 }{81}}\,{e^{-{\frac {20}{3}}\,{\frac {\pi
\,{R}^{3}}{{\rho}^{3}}}}}{R} ^{9}{\pi }^{3}{\rho}^{-9}
\nonumber \\
+{\frac
{20000}{243}}\,{e^{-{\frac {20}{3}}\, {\frac {\pi
\,{R}^{3}}{{\rho}^{3}}}}}{R}^{12}{\pi }^{4}{\rho}^{-12}
\quad .
\label{survivalfn3r}
\end{eqnarray}

\section{Self-similar void distribution}

\label{sectionself}

The statistics of the voids
between galaxies have been analysed
in \cite{Benda-Beckmann2008} with the following self-similar
SF in the following, $S_{SS}$,
\begin{equation}
S_{SS}=
{{\rm e}^{- \left( {\frac {R}{s_{{1}}\lambda}} \right) ^{p_{{1}}}-
 \left( {\frac {R}{s_{{2}}\lambda}} \right) ^{p_{{2}}}}}
\quad  ,
\label{survivalss}
\end{equation}
where $\lambda$ is the mean separation between galaxies,
$s_1$ and $s_2$ are two length factors,
and  $p_1$ and $p_2$ two powers.
The DF of the self-similar distribution
is
\begin{equation}
DF_{SS} =
1-{{\rm e}^{- \left( {\frac {R}{s_{{1}}\lambda}} \right) ^{p_{{1}}}-
 \left( {\frac {R}{s_{{2}}\lambda}} \right) ^{p_{{2}}}}}
\quad .
\label{dfss}
\end{equation}
The PDF of the self-similar distribution
 is
\begin{eqnarray}
p_{SS}(R) = \nonumber  \\
\frac
{
{{\rm e}^{- \left( {\frac {R}{s_{{1}}\lambda}} \right) ^{p_{{1}}}-
 \left( {\frac {R}{s_{{2}}\lambda}} \right) ^{p_{{2}}}}} \left(
 \left( {\frac {R}{s_{{1}}\lambda}} \right) ^{p_{{1}}}p_{{1}}+
 \left( {\frac {R}{s_{{2}}\lambda}} \right) ^{p_{{2}}}p_{{2}}
 \right)
}
{
R
}
\quad .
\end{eqnarray}
At present, it is not possible to find an analytical
expression for the integral that defines the average value
of the self-similar  distribution.

\begin{figure}
\begin{center}
\includegraphics[width=7cm]{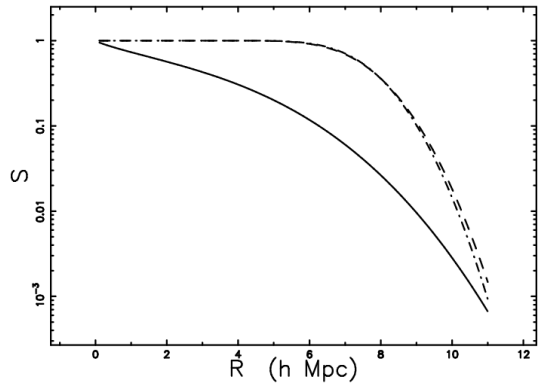}
\end {center}
\caption
{
The survival function, $S_{SS}$,
of the
self-similar distribution in radius of $N/S1$
as represented by (\ref{survivalss})
  (full line),
the survival function, $S_{FN3R}$,
for the Ferenc--Neda function of the radius
 as represented by
(\ref{survivalfn3r})
when $d=3$, $\rho=12.5$~Mpc
and $\chi^2=4319$ with
100 points
 (dashed line). The
survival function, $S_{KR} $,
of the radius of the Kiang function
as represented by
(\ref{survival_kiangr})
when $\rho=12.5$~Mpc, $c=5.53$
and $\chi^2=4076$ with
100 points  (dot-dash-dot-dash line).
}
 \label{comparison_3d}
 \end{figure}

A comparison of the survival function of
self-similar voids with the survival function
of the radii of the two PDFs analysed here is
reported in Figure~\ref{comparison_3d}.

The two fits of Figure~\ref{comparison_3d}
are not satisfactory because we are
making a comparison between
the radius of projected voids
and the 3D radii
of the normalized volume distribution
of the PVT.
This fact is confirmed from the high values
of $\chi^2$ computed according to
(\ref{chisquare}).

\section{The sectional PVT}

\label{sectionsectional}

The existence of voids between galaxies is normally
deduced from a projected distribution
in astronomical slices such as the
2dFGRS~S3, see Figure 1 in \cite{Benda-Beckmann2008}.
This fact motivates the analysis of
the sectional Voronoi tessellation, also known
as $V_p(2,3)$, see \cite{okabe}.
A previous analysis has shown that a cell
belonging to the intersection between
an arbitrary plane and
the faces of the Voronoi
polyhedrons is almost surely  a non-Voronoi cell,
see details in \cite{Weygaert1996}.
Here, we first model the normalized area-distribution
$V_p(2,3)$  with Kiang  PDFs as represented
by  (\ref{kiang}),
see Table~\ref{tablev23}.
 \begin{table}
 \caption[]{
Values of $\chi^2$ for
the cell normalized area-distribution function of $V_p(2,3)$;
here $T_i$ are the  theoretical frequencies and
$O_i$ are the  sample frequencies.
Here we have
8517 Poissonian seeds and 40 intervals in the histogram.
}
 \label{tablev23}
 \[
 \begin{array}{llll}
 \hline
PDF ~  &  parameters & \chi^2  \\ \noalign{\smallskip}
 \hline
 \noalign{\smallskip}
 H (x ;c ) ~(Eq. (\ref{kiang}))
    &  c=2.07 & 114.41   \\
 p (x ; b ) ~(Eq. (\ref{exponential}))
    &  d=1 & 85.38   \\
\noalign{\smallskip}
 \hline
 \hline
 \end{array}
 \]
 \end {table}

In the case of $n$ cuts, we can compute the average value of
$c$, $\overline{c}$, and the standard deviation
$\sigma$, see Table~\ref{averagedc}.

 \begin{table}
 \caption[]{
Average value of $c$ according to the
Kiang function, Eq. (\ref{kiang})),
for $V_p(2,3)$, when 50 cuts are considered.
Here we have
25087 Poissonian seeds.
}
 \label{averagedc}
 \[
 \begin{array}{llll}
 \hline
PDF ~& name  &  \overline{c}   & \sigma   \\ \noalign{\smallskip}
 \hline
 \noalign{\smallskip}
 H (x ;c ) ~(Eq. (\ref{kiang})) & Kiang &
2.25 &  0.11  \\
\noalign{\smallskip}
 \hline
 \hline
 \end{array}
 \]
 \end {table}
A test of this value can be done
on the  unpublished manuscript  of  Ken Brakke
available at
http://www.susqu.edu/brakke/aux/ \newline
downloads/papers/3d.pdf~.
\newline
Table 4 of this paper is dedicated to the
3D plane cross-sectional statistics
E(area) = 0.6858 and
Var(area) = 0.2269.
When  a typical run of ours is normalized to the
average value rather than 1,
our result is Var(area) = 0.2266
which means  that our numerical evaluation differs
by 2/10000 from the theoretical result.
We remember that in the case of the
normalized area-distribution
function in 2D we have  $c \approx 4$
and we can therefore speak of a decrease
in the value of $c$ by a factor 2.
The decrease of $c$ for $V_p(2,3)$
was first derived in Figure~4 of
\cite{Zaninetti2006}.

Another PDF that can be considered in order to
model
the normalized area distribution of $V_p(2,3)$
is the exponential distribution,
\begin{equation}
p(x) = \frac{1} {b} \exp {-\frac{x} {b}}
\label{exponential}
\quad ,
\end{equation}
which has an average value
\begin{equation}
\overline {x} = b
\quad .
\end{equation}
In the case of the normalized areas $b=1$,
Table \ref{tablev23} reports the $\chi^2$ values
of the two distributions adopted here.
Once the statistics  of the normalized
area distribution of an arbitrary cut
$V_p(2,3)$ are known, we can model the radius
distribution.
We therefore convert the area of each cell, $A$,
to an equivalent radius $R$
\begin{equation}
R= \sqrt{ \frac {A} {\pi} }
\quad  .
\end{equation}
We also briefly remember that the problem
of stereology is to deduce
the true size distribution $N(R)$ of 3D-volumes
from the  distribution $n(r)$
of apparent circles in 2D.

\subsection{Kiang distribution of $V_p(2,3)$ in radius }

The PDF, $ H_{R23}  (R ;c )$,
as a function of the radius corresponding
to the Kiang function as represented
by (\ref{kiang}) for
$V_p(2,3)$
 is
\begin{equation}
 H_{R23}  (R ;c )   =
\frac
 {
2\,c \left( {\frac {c\pi \,{R}^{2}}{{\rho}^{2}}} \right) ^{c-1}{
{\rm e}^{-{\frac {c\pi \,{R}^{2}}{{\rho}^{2}}}}}\pi \,R
 }
 {
\Gamma  \left( c \right) {\rho}^{2}
 }
 \quad ,
\label{kiang23}
\end{equation}
where $ 0 \leq R < \infty $, $ c~>0$
and $ \rho ~>0$.
The Kiang  PDF of the radius
for
$V_p(2,3)$
has a mean of
\begin{equation}
\mu =
{\frac {\rho\,\Gamma  \left( c+1/2 \right) }{\sqrt {c}\sqrt {\pi }
\Gamma  \left( c \right) }}
 \quad ,
\end{equation}
and variance
\begin{equation}
\sigma^2 =
\frac
{
{\rho}^{2} \left( c \left( \Gamma  \left( c \right)  \right) ^{2}-
 \left( \Gamma  \left( c+1/2 \right)  \right) ^{2} \right)
}
{
c\pi \, \left( \Gamma  \left( c \right)  \right) ^{2}
}
\quad .
\end{equation}
The  DF of the Kiang function of the radius, $DF_{KR23}$,
for
$V_p(2,3)$ is:
\begin{equation}
DF_{KR23} = 1-
\frac
{
\Gamma  \left( c,2\,{\frac {c\pi \,{R}^{2}}{{\rho}^{2}}} \right)
}
{
\Gamma  \left( c \right)
}
\quad .
\end{equation}

The survival function $S_{KR23}$ of the radius
for
$V_p(2,3)$
 is
\begin{equation}
S_{KR23} =
\frac
{
\Gamma  \left( c,2\,{\frac {c\pi \,{R}^{2}}{{\rho}^{2}}} \right)
}
{
\Gamma  \left( c \right)
}
\quad .
\label{survival_kiangr23}
\end{equation}

\begin{figure}
\begin{center}
\includegraphics[width=7cm]{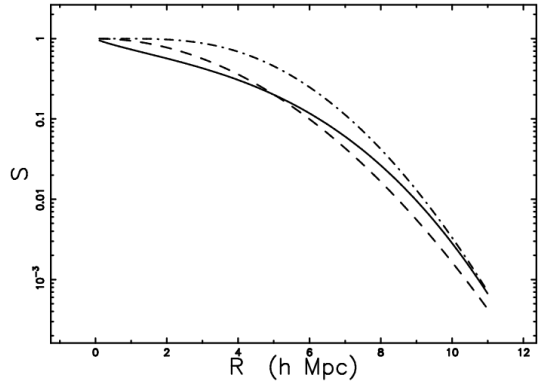}
\end {center}
\caption
{
The survival function, $S_{SS}$,
of the
self-similar  distribution in radius of $N/S1$ as represented by (\ref{survivalss})
  (full line),
the survival function, $S_{KR23} $,
of the radius of the Kiang function for $V_p(2,3)$
as represented by
(\ref{survival_kiangr23})
when $\rho=13$~Mpc, $c=2.25$
and $\chi^2=67.1$ with
100 points
 (dot-dash-dot-dash line).
The survival function, $S_{ER23} $,
of the radius of the exponential distribution
for $V_p(2,3)$
as represented by
(\ref{survival_expr23})
when $\rho=7$~Mpc
and $\chi^2=9.27$ with
100 points
(dashed line).
}
 \label{comparison_cut}%
 \end{figure}

A comparison of the survival
function of self-similar voids with the
survival function
of the radius
for  $V_p(2,3)$
of the exponential
distribution is
reported in Figure~\ref{comparison_cut}.

\subsection{Exponential distribution of $V_p(2,3)$ in radius }

The PDF, $ p_{R23}  (R ;c )$,
as a function of the radius corresponding
to the exponential distribution as represented
by  (\ref{exponential}) for
$V_p(2,3)$
 is
\begin{equation}
 p_{R23}  (R ;c )   =
\frac
 {
2\,{{\rm e}^{-{\frac {\pi \,{R}^{2}}{{\rho}^{2}}}}}\pi \,R
 }
 {
{\rho}^{2}
 }
 \quad ,
\label{exponential23}
\end{equation}
where $ 0 \leq R < \infty $, $ \rho>0$.
The exponential PDF of the radius
for
$V_p(2,3)$
has a mean of
\begin{equation}
\mu =
\frac
{
{\pi }^{3/2}
}
{
2\,{\rho}^{2} \left( {\frac {\pi }{{\rho}^{2}}} \right) ^{3/2}
}
 \quad ,
\label{rmedioexp}
\end{equation}
and variance
\begin{equation}
\sigma^2 =
\frac
{
{\rho}^{2} \left( 4 - \pi  \right)
}
{
4\,\pi
}
\quad .
\end{equation}
The  DF of the exponential distribution
in radius, $DF_{ER23}$,
for
$V_p(2,3)$ is:
\begin{equation}
DF_{ER23} = 1-
{{\rm e}^{-{\frac {\pi \,{R}^{2}}{{\rho}^{2}}}}}
\quad .
\end{equation}

The survival function $S_{ER23}$ of the radius
for
$V_p(2,3)$
 is
\begin{equation}
S_{ER23} =
{{\rm e}^{-{\frac {\pi \,{R}^{2}}{{\rho}^{2}}}}}
\quad .
\label{survival_expr23}
\end{equation}
Figure~\ref{comparison_cut}
reports a
 comparison between the survival
function of self-similar voids
and the exponential  distribution
for $V_p(2,3)$.

In this case, the two types of fit
of Figure ~\ref{comparison_cut}
are satisfactory because
we are making a comparison between
 the observed projected radius of voids
and the projected radius of the Voronoi volumes.
The smaller $\chi^2$ associated with the exponential
distribution can be considered a consequence
of the fact that the statistics
of the normalized area distribution
of the cuts are better described
by an exponential distribution than by the
Kiang function,
see Table~\ref{tablev23}.

\begin{figure}
\begin{center}
\includegraphics[width=7cm]{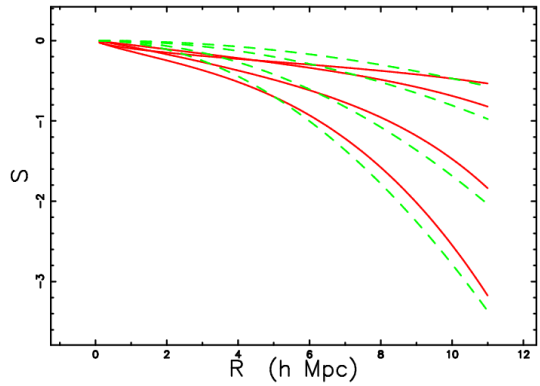}
\end {center}
\caption{
The survival function,
$S_{SS}$,
for the
self-similar distribution in radius of $N/S1$,
$N/S2$, $N/S3$ and $N/S4$,
as reported in Figure 4 of
von Benda-Beckmann and Muller 2008
(full red lines),
as represented by (\ref{survivalss}).
The survival function, $S_{ER23} $,
of the radius of the exponential distribution
for $V_p(2,3)$
as represented by
(\ref{survival_expr23})
when $\rho=7$~Mpc,
     $\rho=9$~Mpc,
     $\rho=13$~Mpc and
     $\rho=17.7$~Mpc
(dashed green  lines).
}
 \label{comparison_sample}%
 \end{figure}

A final comparison between the four samples
of void size statistics as
represented in Figure~4 of
\cite{Benda-Beckmann2008} and our
survival function
of the radius of the exponential distribution
for $V_p(2,3)$
is reported in Figure~\ref{comparison_sample}.

\section{Large-scale structure}

\label{seclarge}
We briefly review  the status of the knowledge
of the Hubble constant,
reference magnitude of the sun,
luminosity function of galaxies,
Malmquist bias and
3D Poissonian Voronoi diagrams.

\subsection{The adopted constants}
A  first
important evaluation of the
Hubble constant is
through Cepheids (key programs with HST) and
type Ia Supernovae, see \cite{Sandage2006},
$
H_0 =(62.3 \pm 5 ) \mathrm{\ km\ s}^{-1}\mathrm{\ Mpc}^{-1}
$.
A second  important evaluation
comes from three years of observations with the
Wilkinson Microwave Anisotropy Probe,
see Table 2 in \cite{Spergel2007},
$
H_{0}=(73.2 \pm 3.2)
\mathrm{\ km\ s}^{-1}\mathrm{\ Mpc}^{-1}
$.
In  the following, we will take the
average value of these two important evaluations,
$H_{0}=67.65 \mathrm{\ km\ s}^{-1}\mathrm{\ Mpc}^{-1} $.
The   Hubble constant  is also reported as \newline
 $H_0 = 100 h \mathrm{\ km\ s}^{-1}\mathrm{\ Mpc}^{-1}$,
with $h=1$
when  $h$ is not specified;  in our  case $h=.6765$.

Another quantity that should be fixed in order
to continue is
the absolute magnitude of
the sun in the $\bj$ filter
of the 2dFGRS
$\mathcal{M_{\sun}}$ = 5.33, see
\cite{Colless2001,Einasto_2009,Eke_2004}.

\subsection{Malmquist bias}

This bias  was originally applied
to the stars,
see \cite{Malmquist_1920, Malmquist_1922},
 and was
then applied to the galaxies by \cite{Behr1951}.
We now introduce the concept of
limiting apparent magnitude and the corresponding
 completeness in
absolute magnitude of the considered catalog
as a function of
redshift.
The observable absolute magnitude as a function of the
limiting apparent magnitude, $m_L$, is
\begin{equation}
M_L =
m_{{L}}-5\,{\it \log_{10}} \left( {\frac {{\it c}\,z}{H_{{0}}}}
 \right) -25
\quad .
\label{absolutel}
\end{equation}
The previous formula predicts, from a theoretical
point of view, an upper limit on the absolute
maximum magnitude that can be observed in a
catalog of galaxies characterized
by a given limiting
magnitude.

The interval covered by the
LF of galaxies,
$\Delta M $,
is defined by
\begin{equation}
\Delta M = M_{max} - M_{min}
\quad ,
\end{equation}
where $M_{max}$ and $M_{min}$ are the
maximum and minimum
absolute
magnitude of the LF for the considered catalog.
The real observable interval in absolute magnitude,
$\Delta M_L $,
 is
\begin{equation}
\Delta M_L = M_{L} - M_{min}
\quad .
\end{equation}
We can therefore introduce the range
of observable absolute maximum magnitude
expressed in percent,
$ \epsilon(z) $,
as
\begin{equation}
\epsilon_s(z) = \frac { \Delta M_L } {\Delta M } \times 100
\, \%
\quad .
\label{range}
\end{equation}
This is a number that represents
the completeness of the sample
and, given the fact that the limiting magnitude of the 2dFGRS is
$m_L$=19.61,
it is possible to conclude that the 2dFGRS is complete
for $z\leq0.0442$~.
This efficiency, expressed as a percentage,
can be considered to be a version  of the Malmquist bias.

\subsection{Luminosity function of galaxies}

The Schechter function, introduced by
\cite{schechter},
provides a useful fit for the
LF (luminosity function) of galaxies
\begin{equation}
\Phi (L) dL = (\frac {\Phi^*}{L^*}) (\frac {L}{L^*})^{\alpha}
\exp \bigl ( {- \frac {L}{L^*}} \bigr ) dL \quad .
\label{equation_schechter}
\end {equation}
Here, $\alpha$ sets the slope
for low values of $L$, $L^*$ is the
characteristic luminosity and $\Phi^*$ is the normalization.
The equivalent distribution in absolute magnitude is
\begin{eqnarray}
\Phi (M)dM=&&(0.4 ln 10) \Phi^* 10^{0.4(\alpha +1 ) (M^*-M)}\nonumber\\
&& \times \exp \bigl ({- 10^{0.4(M^*-M)}} \bigr) dM \quad ,
\label{equation_schechter_M}
\end {eqnarray}
where $M^*$ is the characteristic magnitude as derived from the
data.
The parameters  of the Schechter function
for the  2dFGRS
can be found on the first line of Table~3 in
\cite{Madgwick_2002}
and
are reported in Table~\ref{parameters}.

\begin{table}
 \caption[]{Parameters of the Schechter function for \\
      the 2dFGRS.}
 \label{parameters}
 \[
 \begin{array}{lc}
 \hline
 \hline
 \noalign{\smallskip}
parameter            & 2dFGRS                                  \\ \noalign{\smallskip}
M^* - 5\log_{10}h ~ [mags]         &  ( -19.79 \pm 0.04)           \\ \noalign{\smallskip}
\alpha               &   -1.19  \pm 0.01                       \\ \noalign{\smallskip}
\Phi^* ~[h^3~Mpc^{-3}] &   ((1.59   \pm 0.1)10^{-2})      \\ \noalign{\smallskip}
 \hline
 \hline
 \end{array}
 \]
 \end {table}

We now introduce $f$,
the flux of radiation expressed in units of
$L_{\sun}$~Mpc$^{-2}$,
with $L_{\sun}$ representing the luminosity of the sun.
The joint distribution in {\it z}, redshift
 and {\it f},
see (1.104) in
 \cite{pad}
or (1.117)
in
\cite{Padmanabhan_III_2002},
 is
\begin{equation}
\frac{dN}{d\Omega dz df} =
4 \pi \bigl ( \frac {c_l}{H_0} \bigr )^5 z^4 \Phi (\frac{z^2}{z_{crit}^2})
\label{nfunctionz}
\quad ,
\end {equation}
where $d\Omega$, $dz$, and $ df $
represent the differential of
the solid angle, redshift, flux
and
$c_l$ represents the velocity of light.

This relationship has been derived assuming
$z \approx \frac{V}{c_l} \approx \frac {H_0 r}{c_l}$
with $r$ representing
the distance of the galaxy in Mpc.
The critical value of $z$, $z_{crit}$, is
\begin{equation}
 z_{crit}^2 = \frac {H_0^2 L^* } {4 \pi f c_l^2}
\quad .
\end{equation}
The number of galaxies in
$ z$ and $f$ as given by
(\ref{nfunctionz}) has a maximum at $z=z_{pos-max}$,
where
\begin{equation}
 z_{pos-max} = z_{crit} \sqrt {\alpha +2 }
\quad ,
\end{equation}
which can be re-expressed as
\begin{equation}
 z_{pos-max} =
\frac
{
\sqrt {2+\alpha}\sqrt {{10}^{ 0.4\,{\it M_{\sun}}- 0.4\,{\it M^*}}}{
\it H_0}
}
{
2\,\sqrt {\pi }\sqrt {f}{\it c_l}
}
\quad .
\label{zmax_sch}
\end{equation}

\begin{figure*}
\begin{center}
\includegraphics[width=7cm]{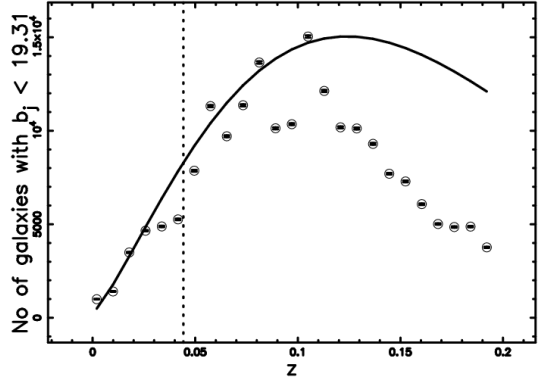}
\end {center}
\caption{
The galaxies in the  2dFGRS     with
$ 13.34 \leq  bJmag \leq 19.61 $
or
 1540  $L_{\sun}$~Mpc$^{-2} \leq
f \leq 493844 L_{\sun}$~Mpc$^{-2}$,
are organized as frequencies versus
heliocentric redshift (empty stars).
The theoretical curve generated by
the integral of the Schechter function in flux,
(\ref{integrale_schechter},
(full line)
is drawn.
The maximum in the frequencies of observed galaxies
is at $z=0.108$.
In this plot, $\mathcal{M_{\sun}}$ = 5.33 and $h$ = 0.67.
The vertical dotted line represents the boundary
between complete and incomplete samples.
}
          \label{maximum_flux_all}%
    \end{figure*}

The number of galaxies, $N_S(z,f_{min},f_{max})$,
comprised between the minimum value of flux,
 $f_{min}$,  and the maximum value of flux, $f_{max}$,
can be computed with the following integral
\begin{equation}
N_S (z) = \int_{f_{min}} ^{f_{max}}
4 \pi  \bigl ( \frac {c_l}{H_0} \bigr )^5    z^4 \Phi (\frac{z^2}{z_{crit}^2})
df
\quad .
\label{integrale_schechter}
\end {equation}
This integral does not  have  an analytical solution
and therefore
a numerical integration must be performed.
The total number of galaxies in the 2dFGRS
is reported in Figure~\ref{maximum_flux_all} as well
as the theoretical curves as represented
by the numerical integration of (\ref{nfunctionz}).
A careful inspection of the previous figure   allows  to conclude
that the theoretical integral fits well the experimental data up
to $z=0.0442$. Beyond that value, the presence of the Malmquist
bias decreases the number of observable galaxies and the
comparison between theory and observations cannot be made.

\subsection{Voronoi diagrams}

\begin{figure*}
\begin{center}
\includegraphics[width=7cm]{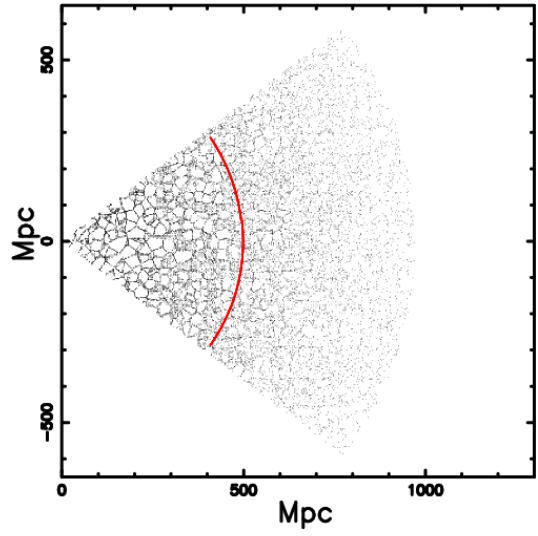}
\end {center}
\caption{
A Voronoi slice
$75^{\circ}$ long and  $3^{\circ}$
wide.
The range of the flux is  2500  $L_{\sun}$~Mpc$^{-2} \leq
f \leq 286808  L_{\sun}$~Mpc$^{-2}$
and the number of galaxies is 19895.
The red circle denotes the confusion distance as
given by (\ref{confusion}).
}
          \label{una_voronoi}%
    \end{figure*}

Voronoi diagrams represent a useful tool for describing the
spatial distribution of galaxies  and, as an example, \cite
{Weygaert1989} identified the  vertexes of irregular Voronoi
polyhedrons with Abell clusters. Another example is provided by
\cite{zaninettig} where the galaxies were first inserted on the
faces of the irregular Voronoi polyhedrons and a power law
distribution for the seeds was adopted. Later, the galaxies were
still inserted on the faces of the irregular Voronoi polyhedrons
but Poissonian seeds were adopted, see
\cite{Zaninetti2006,Zaninetti2008b,Zaninetti2010a}. Adopting the
same algorithm developed in \cite{Zaninetti2010a}, one slice of
the 2dFGRS with the number of galaxies as a function of $z$ as
given by (\ref{integrale_schechter}) is simulated, see
Figure~\ref{una_voronoi}.

This simulation should be compared with Figure~1 in
\cite{Benda-Beckmann2008} and
with  a $3^{\circ}$ slice  of the
2dFGRS Image Gallery
available at the web site, http://msowww.anu.edu.au/2dFGRS/.
Little need be said about the number of seeds
which  should be used in order to simulate the observed slices
of the 2dFGRS.
The average volume of the voids
is the side of the box in Mpc  divided by
the number of seeds.
The average diameter of the voids, $D_{Voronoi}$,
is
\begin{equation}
D_{Voronoi} = ( \frac {side [Mpc]}{No~of~seeds})^{1/3}
\quad .
\end{equation}
The theoretical average diameter, $D_{th}$,
can be obtained from
the average value (\ref{rmedioexp} ) of the
exponential  distribution of $V_p(2,3)$ in radius
once the  maximum value
of $\rho$ which fits
the \cite{Benda-Beckmann2008} data,
see Figure~\ref{comparison_sample},  is adopted
\begin{equation}
D_{th} = \mu \frac{2}{h}
\quad .
\end{equation}
The equality $D_{Voronoi}=D_{th} $ allows to obtain
 the number of
Poissonian seeds.

Particular attention should be paid to the fact that
the astronomical slices are not a plane which intersects
a Voronoi network.
In order to quantify this effect,
we introduce a confusion
distance, $D_c$, as the distance  after which
the altitude of the slices equalizes the
observed average diameter  $\overline{D_{obs}}$
\begin{equation}
D_c \tan (\alpha) =  \overline{D_{obs}}
\quad  ,
\label{confusion}
\end{equation}
where $\alpha$  is the opening angle  of the slice
and $\overline{D_{obs}}$ is the average diameter of the
voids.
In the case of  2dFGRS   $\alpha=3^{\circ}$
and therefore $D_c =498.5$~Mpc
when $D_{obs}= 26.12$~Mpc, see the red circle in
Figure \ref{una_voronoi}.

\section{Conclusions}
The  PDFs  which are usually used to model the
normalized volume distribution of the 3D
PVT are
gamma type distributions such  as  the
Kiang function, (\ref{kiang}),
and the new
Ferenc--Neda function, (\ref{rumeni}).

Here, in order to make a comparison
with the self-similar distribution of voids,
we derived the survival distribution
of the Kiang function, (\ref{survival_kiang}),
and of the Ferenc--Neda function,
 (\ref{survivalfn3}).

On adopting an  astrophysical point of view  the cut  $V_p(2,3)$
may model the voids between galaxies as given by astronomical
slices of  the Millenium catalog. The analysis of the normalized
areas of $V_p(2,3)$ is a subject of research rather than a
well-established fact and we have fitted them with the Kiang
function and the exponential distribution. The $\chi^2$ value
indicates that the exponential distribution fits more closely the
normalized area distribution of $V_p(2,3)$ than does the Kiang
function, see Table~\ref{tablev23}. This fact follows from the
comparison between the self-similar survival function and the
exponential and Kiang distributions of the radius, see
Figure~\ref{comparison_cut}. Therefore, the one parameter survival
function of the radius of the exponential distribution for
$V_p(2,3)$, $S_{ER23} $, as represented by
(\ref{survival_expr23}), may model the voids between galaxies as
well as the five parameter self-similar survival function.

The  observed 2dFGRS slices can be simulated
but the behaviour of the luminosity function for
galaxies and the consequent number of galaxies
as a function of the redshift should be carefully
analysed.

We are planning,
in future projects,
\begin{itemize}
\item To insert the thickness
of the faces of PVT
and to model the connected modification of the
survival function.
\item
To  apply  the
technique  here developed
to the real data sets, Millennium simulation,
for example. The 3D nature of
the method and detailed density mapping of
the voids should be superior to other void
identification algorithms as suggested
by \cite{Schaap2009}.
\end{itemize}


\end{document}